\documentclass[10pt]{IEEEtran}
\newif\ifpdf\ifx\pdfoutput\undefined\pdffalse\else\pdfoutput=1\pdftrue\fi
\usepackage[pdftex]{graphicx}
\usepackage{amssymb, amsmath, cite}
\usepackage{graphicx,subfigure}

\title{On the Capacity of Pairwise Collaborative Networks}

\author{\authorblockN{Saeed~A.~Astaneh$^{^{\dagger}}$, Saeed~Gazor$^{^{\dagger}}$,
Hamid~Behroozi$^{^{\dagger}{}^{\dagger}}$\\
\authorblockA{$^{^{\dagger}}$Department of Electrical and Computer Engineering, Queen's University, Kingston, ON, Canada \\
$^{^{\dagger}{}^{\dagger}}$Department of Mathematics and Statistics,
Queen's University, Kingston, ON, Canada\\
Email: astaneh, s.gazor, behroozi@queensu.ca}}}

\begin{document}

\maketitle

\begin{abstract}
We derive expressions for the achievable rate region of a collaborative coding scheme in a two-transmitter, two-receiver Pairwise Collaborative Network (PCN) where one transmitter and receiver pair, namely relay pair, assists the other pair, namely the source pair, by partially decoding and forwarding the transmitted message to the intended receiver. The relay pair provides such assistance while handling a private message. We assume that users can use the past channel outputs and can transmit and receive at the same time and in the same frequency band. In this collaborative scheme, the transmitter of the source pair splits its information into two independent parts. Ironically, the relay pair employs the decode and forward coding to assist the source pair in delivering a part of its message and re-encodes the decoded message along with private message, which is intended to the receiver of the relay pair, and broadcasts the results. The receiver of the relay pair decodes both messages, retrieves the private message, re-encodes and transmits the decoded massage to the intended destination. We also characterize the achievable rate region for Gaussian PCN. Finally, we provide numerical results to study the rate trade off for the involved pairs. Numerical result shows that the collaboration offers gain when the channel gain between the users of the relay pair are strong. It also shows that if the channel conditions between transmitters or between the receivers of the relay and source pairs are poor, such a collaboration is not beneficial.

\end{abstract}

\section{Introduction}
\label{sec:intro}
\begin{figure}
\begin{center}
\includegraphics[width=\columnwidth]{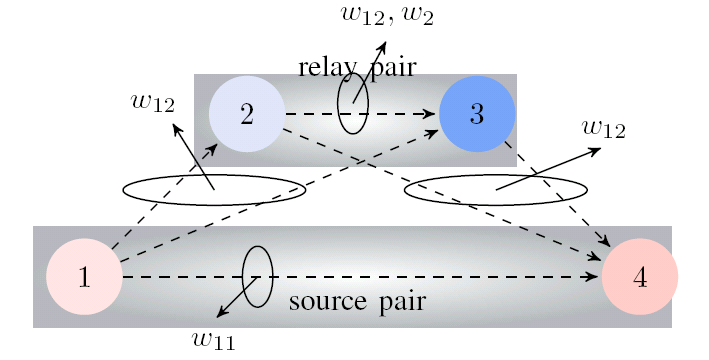}
\caption{Collaboration schemes in the PCN: the partial decode and forward scheme; the source pair transmitter splits its information into two independent parts. The relay pair employs decode and forward scheme to assist the source pair in delivering only one part of its message. 
}
\label{fig:nettopology}
\end{center}
\end{figure}

In a multi user network users may collaborate to jointly convey the information. Van der Meulen \cite{vandermeulen1971ttc} introduced the relay channel where a relay forwards the data from a source to the destination. Cover and El Gamal \cite{cover1979ctr} proved some capacity theorems for a single relay channel. In a collaborative network, users may collaborate to transmit message of other users while handling their own private messages; this can be regarded as a generalization of the traditional relay channel. 
We present a pairwise relaying collaboration model where a pair of transmitter and receiver collaborates with the source pair in delivering the message of the source pair along with its own private message. Our proposed model differs from previous research in that we consider collaboration schemes that the transmitter and receiver of the relay pair handles a private message, which, to the best of our knowledge, no previous work has considered in this setting. Figure~\ref{fig:nettopology} represents such a network where the 1st user, intends to send a message to the 4th user, and the 2nd user to the 3rd user. We propose two collaboration schemes where in the first scheme, the transmitter of the relay pair, the 1st user, splits its message into two independent parts. The relay pair collaborates with the source pair via decode and forward coding to transmit a part of the message of the source pair and the private message of the relay pair. In the second scheme the relay pair partially cancels the interference of other users and sends the compressed observed signal to the intended receiver of source pair, the 4th user.

Collaboration between wireless users has been investigated recently by several authors. Liang and Veeravalli \cite{liang2006crb} studied a cooperative relay broadcast channel with three users where relay links are incorporated into standard two-user broadcast channels to support user cooperation. Liang and Kramer \cite{liang2007rrr} have found improved bounds for the relay broadcast channel. Tannious and Nosratinia in \cite{tannious2007rcp} developed decode and forward and compress and forward strategies for a network of one relay channel with private messages where in addition to the traditional communication from source to destination (assisted by relay), the source has a private message for the relay, and the relay has a private message for the destination (see \cite{kramer2005csa} for a survey on decode and forward and compress and forward strategies). Akhavan and Gazor \cite{akhavanastaneh2007cgr} investigated multi-hopping strategies and resource allocation in such networks. Reznik, Kulkarni and Verdu in \cite{reznik2005brc} further studied the relay broadcast collaborative model for the case of more than two destinations. Sendonaris, Erkip and Aazhang in \cite{sendonaris2003ucd1, sendonaris2003ucd2} showed that collaboration enlarges the achievable rate region in a channel with two collaborative transmitters and a single receiver. Laneman, Tse and Wornell considered a fading channel with two cooperative transmitters and two non-cooperative receivers \cite{laneman2004cdw}. Host-Madsen in \cite{hostmadsen2004arr, hostmadsen2006cbc} presented the achievable rate regions for channels with transmitter and/or receiver collaboration. Ng, Jindal, Goldsmith and Mitra \cite{ng2007cgt} investigated capacity improvement from transmitter and receiver cooperation in a two-transmitter, two-receiver network.

In this paper, we extend the results of \cite{tannious2007rcp, akhavanastaneh2007cgr} and study the achievable rate region of the decode and forward
coding schemes in the PCN. We present in Section~\ref{sec:mod} the network model. In Section~\ref{sec:rsdf} we develop the rate splitting in conjunction with decode and forward coding scheme for the PCN and determine its capacity. 
We investigate the additive white Gaussian noise PCN rate region in Section~\ref{sec:gc}. Finally in Section~\ref{sec:conclusion} we give the concluding remarks.

\section{System Model}
\label{sec:mod}
The PCN consists of inputs $x_{i}$ where $i \in\left\{ {1,2,3} \right\}$, outputs $y_{j}$ where $j \in \left\{ {2,3,4} \right\}$ and the transition probability $p\left( {y_2 ,y_3 ,y_4 \left| {x_1 ,x_2 ,x_3} \right. } \right)$ (see Figure~\ref{fig:nettopology}). The 1st user wishes to send the message $w_{1}$ to the 4th user, while the 2nd user wishes to send the message $w_{2}$ to the 3rd user.

We define an additive white Gaussian noise (AWGN) PCN with the input output relation:
$  Y_i = Z_i + \sum\limits_{j \ne i} {\sqrt{h_{ij}} X_j }$ where $X_i$, $Y_i$ and $Z_i$ denote input, output and channel noise with normal distributions, i.e. $Z_i \sim \mathcal{N}\left( {0,N_i } \right) $, respectively. Let denote the power gain of the communication channel between the $i$th and $j$th user by $h_{ij}$. We impose the power constraints $E\left( {X_i^2 } \right) \leqslant P_i$ for all channel inputs. We assume that the users can transmit and receive at the same time and in the same frequency band.

In this paper, let $X$, $x$ and $\underline{x}$ denote a random variable, a scalar and a vector, respectively. We define $\bar x = 1 - x$ and $\mathcal{C}(x) = \frac{1}{2} \log(1+x)$.
\section{Collaboration via Partial Decode-and-Forward}
\label{sec:rsdf}
In this section we consider a collaborative scheme, partial decode and forward, which includes rate splitting technique at the source pair transmitter and decode and forward relaying at the relay pair transmitter and receiver. The source pair transmitter splits the message $w_{1}$ into two independent parts, $w_{11}$ and $w_{12}$. 
The source pair transmitter, 1st user, encodes $w_{11}$ and $w_{12}$ to the codeword $x_{1}$. The relay pair transmitter, 2nd user, decodes $w_{12}$ and re-encodes both its message, i.e. $w_{2}$, and $w_{12}$, to $x_{2}$. The relay pair receiver, 3rd user, retrieves $w_{12}$ and $w_2$ from $y_{3}$ and re-encodes $w_{12}$ to $x_{3}$. Finally, the source pair receiver, 4th user, by using $y_{4}$ estimates its intended message $w_{12}$ and $w_{11}$.

In the following we prove that the rates $(R_{1},R_{2})$, given by (\ref{equ:dfcap1}) shown at the top of the next page, are achievable for the PCN for some joint distribution

$ p\left( {x_3 } \right) p\left( {u_1 \left| {x_3 } \right.} \right) p\left( {u_2 \left| {u_1 x_3 } \right.} \right) p\left( {x_2 \left| {u_1 x_3 } \right.} \right) p\left( {x_1 \left| {u_1 u_2 x_3 } \right.} \right)$.

\begin{figure*}
\begin{eqnarray}
\left\{
\begin{gathered}
\label{equ:dfcap1}
  R_{11}  < \min \left\{ {I\left( {Y_4 ;X_1 \left| {U_1 ,U_2 ,X_2 ,X_3 } \right.} \right),I\left( {Y_3 ;X_1 \left| {X_2 ,X_3 ,U_1 ,U_2 } \right.} \right)} \right\} \hfill \\
  R_{12}  < I\left( {Y_2 ;U_2 \left| {U_1 ,X_2 ,X_3 } \right.} \right) \hfill \\
  R_2  < \min \left\{ {I\left( {Y_3 ;X_2 \left| {X_1 ,X_3 ,U_1 ,U_2 } \right.} \right),I\left( {Y_4 ;X_2 \left| {U_1 ,U_2 ,X_1 ,X_3 } \right.} \right)} \right\} \hfill \\
  R_2  + R_{11}  < \min \left\{ {I\left( {Y_3 ;X_1 ,X_2 \left| {X_3 ,U_1 ,U_2 } \right.} \right),I\left( {Y_4 ;X_1 ,X_2 \left| {U_1 ,U_2 ,X_3 } \right.} \right)} \right\} \hfill \\
  R_2  + R_{11}  + R_{12}  < \min \left\{ {I\left( {Y_3 ;X_1 ,X_2 ,U_1 ,U_2 \left| {X_3 } \right.} \right),I\left( {Y_4 ;U_1 ,U_2 ,X_1 ,X_2 ,X_3 } \right)} \right\} \hfill \\
  R_1  = R_{11}  + R_{12}  \hfill \\
\end{gathered}\right.
\hfill \\
\left\{
\begin{gathered}
\label{equ:dfcap2}
  R_1  < I\left( {Y_2 ;U_2 \left| {U_1 ,X_2 ,X_3 } \right.} \right) + \phi _1  \hfill \\
  R_2  < \min \left\{ {I\left( {Y_3 ;X_2 \left| {U_1 ,U_2 ,X_1 ,X_3 } \right.} \right),I\left( {Y_4 ;X_2 \left| {U_1 ,U_2 ,X_1 ,X_3 } \right.} \right)} \right\} \hfill \\
  R_1  + R_2  < \min \left\{ {\phi _2 ,I\left( {Y_2 ;U_2 \left| {U_1 ,X_2 ,X_3 } \right.} \right) + \phi _3 } \right\} \hfill \\
  \phi _1  = \min \left\{ {I\left( {Y_3 ;X_1 \left| {U_1 ,U_2 ,X_2 ,X_3 } \right.} \right),I\left( {Y_4 ;X_1 \left| {U_1 ,U_2 ,X_2 ,X_3 } \right.} \right)} \right\} \hfill \\
  \phi _2  = \min \left\{ {I\left( {Y_4 ;U_1 ,U_2 ,X_1 ,X_2 ,X_3 } \right),I\left( {Y_3 ;U_1 ,U_2 ,X_1 ,X_2 \left| {X_3 } \right.} \right)} \right\} \hfill \\
  \phi _3  = \min \left\{ {I\left( {Y_3 ;X_1 ,X_2 \left| {U_1 ,U_2 ,X_3 } \right.} \right),I\left( {Y_4 ;X_1 ,X_2 \left| {U_1 ,U_2 ,X_3 } \right.} \right)} \right\} \hfill \\
\end{gathered}\right.
\end{eqnarray}
\end{figure*}

For this achievable region, we apply Fourier-Motzkin elimination to eliminate $R_{11}$ and $R_{12}$ from the bounds and then obtain the region (\ref{equ:dfcap2}) which provides a simpler form.

We use the coding strategies developed in \cite{cover1979ctr, kramer2005csa, tannious2007rcp, cover1980mac} for relay and multiple access channels (MACs). The 1st user uses a three-level superposition block Markov encoding, while the 2nd user uses a two-level and the 3rd user a single-level superposition coding. Furthermore, we use the regular encoding/backward decoding techniques. We divide the messages $w_1$ and $w_2$ into $B$ blocks for $b=1,2, ..., B$ and send these message blocks in $B+2$ transmission blocks. In the following we construct the codebooks and discuss the decoding in each block.

\emph{Random Codebook Construction:}

\begin{enumerate}
\item We generate $2^{n R_{12}}$ i.i.d. $\underline{x}_3 =(x_{31}, x_{32}, ... ,x_{3n})$ sequences, each with distribution $p\left( \underline{x}_3 \right) = \prod\limits_{i = 1}^n {p\left( {x_{3i} } \right)}$ and label them $ \underline{x}_3\left( {w''_{12} } \right)$.
\item For each $\underline{x}_{3}\left( {w''_{12} } \right)$, we generate $2^{n R_{12}}$ i.i.d. $\underline{u}_{1}$ sequences, each with distribution $p\left( \underline{u}_{1} \right) = \prod\limits_{i = 1}^n {p\left( {u_{1i} \left| {x_{3i} } \right.} \right)}$ and label them $ \underline{u}_{1}\left({w'_{12} ,w''_{12} } \right)$.
\item For each pair $\underline{u}_{1}\left( {w'_{12} ,w''_{12} } \right)$ and $ \underline{x}_3\left( {w''_{12} } \right)$, we generate $2^{n R_{12}}$ i.i.d. $\underline{u}_2$ sequences, each with distribution $p\left( \underline{u}_{2} \right) = \prod\limits_{i = 1}^n {p\left( {u_{2i} \left| {x_{3i}, u_{1i} } \right.} \right)}$ and label them $ \underline{u}_2\left( {w_{12}, w'_{12} ,w''_{12} } \right)$.
\item For each pair $\underline{u}_{1}\left( {w'_{12} ,w''_{12} } \right)$ and $ \underline{x}_3\left( {w''_{12} } \right)$, we generate $2^{n R_{2}}$ i.i.d. $\underline{x}_2$ sequences, each with distribution $p\left( \underline{x}_{2} \right) = \prod\limits_{i = 1}^n {p\left( {x_{2i} \left| {x_{3i}, u_{1i} } \right.} \right)}$ and label them $ \underline{x}_2\left( {w_{2}, w'_{12} ,w''_{12} } \right)$.
\item For each triplet $\underline{u}_{1}\left( {w'_{12} ,w''_{12} } \right)$, $ \underline{x}_3\left( {w''_{12} } \right)$ and $ \underline{u}_2\left( {w_{12}, w'_{12} ,w''_{12} } \right)$, we generate $2^{n R_{11}}$ i.i.d. $\underline{x}_1$ sequences, each with distribution $p\left( \underline{x}_{1} \right) = \prod\limits_{i = 1}^n {p\left( {x_{1i} \left| {x_{3i}, u_{1i}, u_{2i} } \right.} \right)}$ and label them $ \underline{x}_1\left( {w_{11}, w_{12}, w'_{12} ,w''_{12} } \right)$.
\end{enumerate}

\emph{Encoding:}

For each time $b = 1, 2, ..., B + 2$ the users send the following sequences:
\begin{enumerate}
   \item $\underline{x}_1 \left( {w_{11,\rm b}, w_{12,\rm b},1 ,1 } \right)$, $\underline{x}_2 \left( {w_{2,\rm b} ,1 ,1 } \right)$, $\underline{x}_3 \left( {1} \right)$ $b=1$
   \item $\underline{x}_1 \left( {w_{11,\rm b}}, {w_{12,\rm b}, w_{12,\rm b-1} ,1 } \right)$, $\underline{x}_2 \left( {w_{2,\rm b} ,w_{12,\rm b-1} ,1 } \right)$, $\underline{x}_3 \left( {1} \right)$ $b=2$
   \item $\underline{x}_1 \left( {w_{11,\rm b}, w_{12,\rm b} ,w_{12,\rm b-1} ,w_{12,\rm b-2} } \right)$, $\underline{x}_2 \left( {w_{2,\rm b} ,w_{12,\rm b-1} ,w_{12,\rm b-2} } \right)$, $\underline{x}_3 \left( {w_{12,\rm b-2}} \right)$ $b=3,..., B$
   \item $\underline{x}_1 \left( {1, 1 ,w_{12,\rm b-1} ,w_{12,\rm b-2} } \right)$, $\underline{x}_2 \left( {1 ,w_{12,\rm b-1} ,w_{12,\rm b-2} } \right)$, $\underline{x}_3 \left( {w_{12,\rm b-2}} \right)$, $b=B+1$
   \item $\underline{x}_1 \left( {1, 1 ,1 ,w_{12,\rm b-2} } \right)$, $\underline{x}_2 \left( {1 ,1 ,w_{12,\rm b-2} } \right)$, $\underline{x}_3 \left( {w_{12,\rm b-2}} \right)$ $b=B+2$
\end{enumerate}

\emph{Decoding:}

\begin{enumerate}
  \item The 2nd user decodes $w_{12,\rm b}$ by looking for $\hat w_{12,\rm b}$ such that $\underline{y}_{2, \rm b}$, $ \underline{x}_2\left( {w_{2,\rm b}, w_{12,\rm b-1} ,w_{12,\rm b-2} } \right)$, $ \underline{x}_3\left( {w_{12,\rm b-2} } \right)$, $ \underline{u}_{1}\left({w_{12,\rm b-1} ,w_{12,\rm b-2} } \right)$ and $ \underline{u}_{2}\left( {w_{12, \rm b}, w_{12,\rm b-1} ,w_{12,\rm b-2} } \right)$ are jointly typical. The decoding is reliable if
    $ R_{12} < I\left( {Y_2 ;U_2 \left| {X_2 ,X_3 ,U_1} \right.} \right). $
  \item The 3rd user decodes $w_{12,\rm b}$, $w_{12,\rm b-1}$, $w_{11,\rm b}$ and $w_{2,\rm b}$ by looking for $\hat w_{12,\rm b}$, $\hat w_{12,\rm b-1}$, $\hat w_{11,\rm b}$ and $\hat w_{2,\rm b}$ such that  $\underline{y}_{3, \rm b}$, $ \underline{x}_{1}\left( {w_{11,\rm b}, w_{12,\rm b}, w_{12,\rm b-1} ,w_{12,\rm b-2} } \right)$, $ \underline{x}_{2}\left( {w_{2,\rm b}, w_{12,\rm b-1} ,w_{12,\rm b-2} } \right)$, $ \underline{x}_{3}\left( {w_{12,\rm b-2} } \right)$, $ \underline{u}_{1}\left({w_{12,\rm b-1} ,w_{12,\rm b-2} } \right)$ and $ \underline{u}_{2}\left( {w_{12, \rm b}, w_{12,\rm b-1} ,w_{12,\rm b-2} } \right)$  are jointly typical. Here, the 1st and 2nd users attempt to transmit a common message $w_{12}$ along with their private messages, i.e. $w_{11}$ and $w_{2}$, respectively. It is shown in \cite{cover1980mac, liang2008mac} that this step can be made reliably if \[ \left\{ \begin{gathered}
          R_2  < I\left( {Y_3 ;X_2 \left| {X_1 ,X_3 ,U_1 ,U_2 } \right.} \right) \hfill \\
          R_{11}  < I\left( {Y_3 ;X_1 \left| {X_2 ,X_3 ,U_1 ,U_2 } \right.} \right) \hfill \\
          R_2  + R_{11}  < I\left( {Y_3 ;X_1 ,X_2 \left| {X_3 ,U_1 ,U_2 } \right.} \right) \hfill \\
          R_2  + R_{11}  + R_{12}  < I\left( {Y_3 ;X_1 ,X_2 ,U_1 ,U_2 \left| {X_3 } \right.} \right) \hfill \\
        \end{gathered}  \right.\]
   \item The 4th user decodes $w_{2,\rm b}$, $w_{12,\rm b}$ and $w_{12,\rm b-1}$ and $w_{12,\rm b-2}$. We have the problem of sending correlated sources over a MAC as treated in \cite{cover1980mac, liang2008mac}. The decoding can be done reliably if \[
\left\{ \begin{gathered}
  R_{11}  < I\left( {Y_4 ;X_1 \left| {U_1 ,U_2 ,X_2 ,X_3 } \right.} \right) \hfill \\
  R_2  < I\left( {Y_4 ;X_2 \left| {U_1 ,U_2 ,X_1 ,X_3 } \right.} \right) \hfill \\
  R_{11}  + R_2  < I\left( {Y_4 ;X_1 ,X_2 \left| {U_1 ,U_2 ,X_3 } \right.} \right) \hfill \\
  R_{11}  + R_{12}  + R_2  < I\left( {Y_4 ;U_1 ,U_2 ,X_1 ,X_2 ,X_3 } \right) \hfill \\
\end{gathered}  \right.
\]
 \end{enumerate}
The rate region in (\ref{equ:dfcap1}) follows from combining the achievable regions derived above.

\begin{figure*}

\begin{eqnarray}
\label{equ:dfgas}
\left\{ \begin{gathered}
  \phi _1  = \min \left\{ {\mathcal{C}\left( {\frac{{\bar \alpha h_{13} P_1 }}
{{N_3 }}} \right),\mathcal{C}\left( {\frac{{\bar \alpha h_{14} P_1 }}
{{N_4 }}} \right)} \right\} \hfill \\
  \phi _2  = \min \left\{ \begin{gathered}
  \mathcal{C}\left( {h_{14} P_1 \frac{{\bar \alpha  + \alpha \bar \beta  + \alpha \beta \bar \gamma \left( {1 + \sqrt {\frac{{h_{24} }}
{{h_{14} }}\frac{{\delta P_2 }}
{{\alpha \beta P_1 }}} } \right)^2  + \alpha \beta \gamma \left( {1 + \sqrt {\frac{{h_{24} }}
{{h_{14} }}\frac{{\delta P_2 }}
{{\alpha \beta P_1 }}}  + \sqrt {\frac{{h_{34} }}
{{h_{14} }}\frac{{P_3 }}
{{\alpha \beta \gamma P_1 }}} } \right)^2 }}
{{N_4 }}} \right), \hfill \\
  \mathcal{C}\left( {h_{13} P_1 \frac{{\bar \alpha  + \alpha \bar \beta  + \alpha \beta \bar \gamma \left( {1 + \sqrt {\frac{{h_{23} }}
{{h_{13} }}\frac{{\delta P_2 }}
{{\alpha \beta P_1 }}} } \right)^2  + \frac{{h_{23} }}
{{h_{13} }}\frac{{\bar \delta P_2 }}
{{P_1 }}}}
{{N_3 }}} \right) \hfill \\
\end{gathered}  \right\} \hfill \\
  \phi _3  = \min \left\{ {\mathcal{C}\left( {\frac{{\bar \alpha h_{13} P_1  + \bar \delta h_{23} P_2 }}
{{N_3 }}} \right),\mathcal{C}\left( {\frac{{\bar \alpha h_{14} P_1  + \bar \delta h_{24} P_2 }}
{{N_4 }}} \right)} \right\} \hfill \\
\end{gathered}  \right.
\end{eqnarray}
\end{figure*}

\section{Capacities of AWGN PCNs}
\label{sec:gc}
In this section, we investigate the achievable rate region of involved pairs when two pairs of source and relay collaborate in sending information to the intended receivers. We compare the achievable rate of the proposed collaboration scheme with the scenario where pairs do not collaborate. In the absence of collaboration between pairs, we model the channel by interference channel in which transmission of information of a pair interferes with the communication between the other pair \cite{cover2006eit, sason2004arr}.

The capacity of the interference channel (IFC) is an open problem, however, in the case where $h_{14} \le h_{24}$ and $h_{23} \le h_{13}$ the capacity region of interference channel is completely characterized as:

\begin{equation}
\left\{ {\begin{array}{*{20}l}
   {R_1 } &  <  & {\mathcal{C}\left( {\frac{{h_{14} P_1 }}
{{N_4 }}} \right)}  \\
   {R_2 } &  <  & {\mathcal{C}\left( {\frac{{h_{23} P_2 }}
{{N_3 }}} \right)}  \\
   {R_1  + R_2 } &  <  & {\min \left\{ {\mathcal{C}\left( {\frac{{h_{14} P_1  + h_{24} P_2 }}
{{N_4 }}} \right),\mathcal{C}\left( {\frac{{h_{23} P_2  + h_{13} P_1 }}
{{N_3 }}} \right)} \right\}}  \\
 \end{array} } \right.
\end{equation}

Now, we concentrate on the AWGN PCN. Employing partial decode and forward scheme, the rates $(R_{1},R_{2})$, are achievable for the AWGN PCN:
\begin{eqnarray}
\left\{
\begin{gathered}
  R_1  < \mathcal{C}\left( {\frac{{\alpha \bar \beta h_{12} P_1 }}
{{\bar \alpha h_{12} P_1  + N_2 }}} \right) + \phi _1  \hfill \\
  R_2  < \min \left\{ {\mathcal{C}\left( {\frac{{\bar  \delta h_{23} P_2 }}
{{N_3 }}} \right),\mathcal{C}\left( {\frac{{\bar \delta h_{24} P_2 }}
{{N_4 }}} \right)} \right\} \hfill \\
  R_1  + R_2  < \min \left\{ {\phi _2 ,\mathcal{C}\left( {\frac{{\alpha \bar \beta h_{12} P_1 }}
{{\bar  \alpha h_{12} P_1  + N_2 }}} \right) + \phi _3 } \right\} \hfill \\
\end{gathered}
\right.
\end{eqnarray}
where $\phi _1, \phi _2$ and $\phi _3$ are given by (\ref{equ:dfgas}) shown at the top of the following page.
We use the following independent normal distributions to find the rate region for the partial decode and forward coding scheme (\ref{equ:dfcap2}):
$A \sim \mathcal{N}\left( {0,\bar \alpha P_1 } \right),\rm B \sim \mathcal{N}\left( {0,\alpha \bar \beta P_1 } \right),C \sim \mathcal{N}\left( {0,\alpha \beta \bar \gamma P_1 } \right),D \sim \mathcal{N}\left( {0,\alpha \beta \gamma P_1 } \right),E \sim \mathcal{N}\left( {0,\bar \delta P_3 } \right)$, where $\alpha, \beta, \gamma, \delta \in [0,1]$. Furthermore, we let $X_1 = A + B + C + D$, $ X_2 = E + C + D $, $
X_3 = D$, $U_{1} = C + D$ and $U_{2} = B + C + D$. 

Here, we move on to study the condition under which collaboration improves the achievable rate of pairs. Numerical result shows that the collaboration offers capacity gain when the channel gain between the source pair is week and the corresponding channel between relay users is strong. It also shows that if the channels condition between the transmitters, $h_{12}$ and between the receivers, $h_{34}$, are poor, such a collaboration is not beneficial.

We consider the AWGN PCN with $P_1 = P_2 = P_3 = 1$ and $N_2 = N_3 = N_4 = 1$ and we examine the proposed collaboration scheme under different channel conditions. We compare the achievable rate region of the proposed scheme with the scheme where pairs do not collaborate, i.e. interference channel. We also consider the scenario where both pairs acquire the same rate as $R_1 = R_2$ and study the capacity gain of the schemes.

\begin{figure}
\begin{center}
\subfigure[]{\label{fig:rate1}\includegraphics[width=0.5\columnwidth]{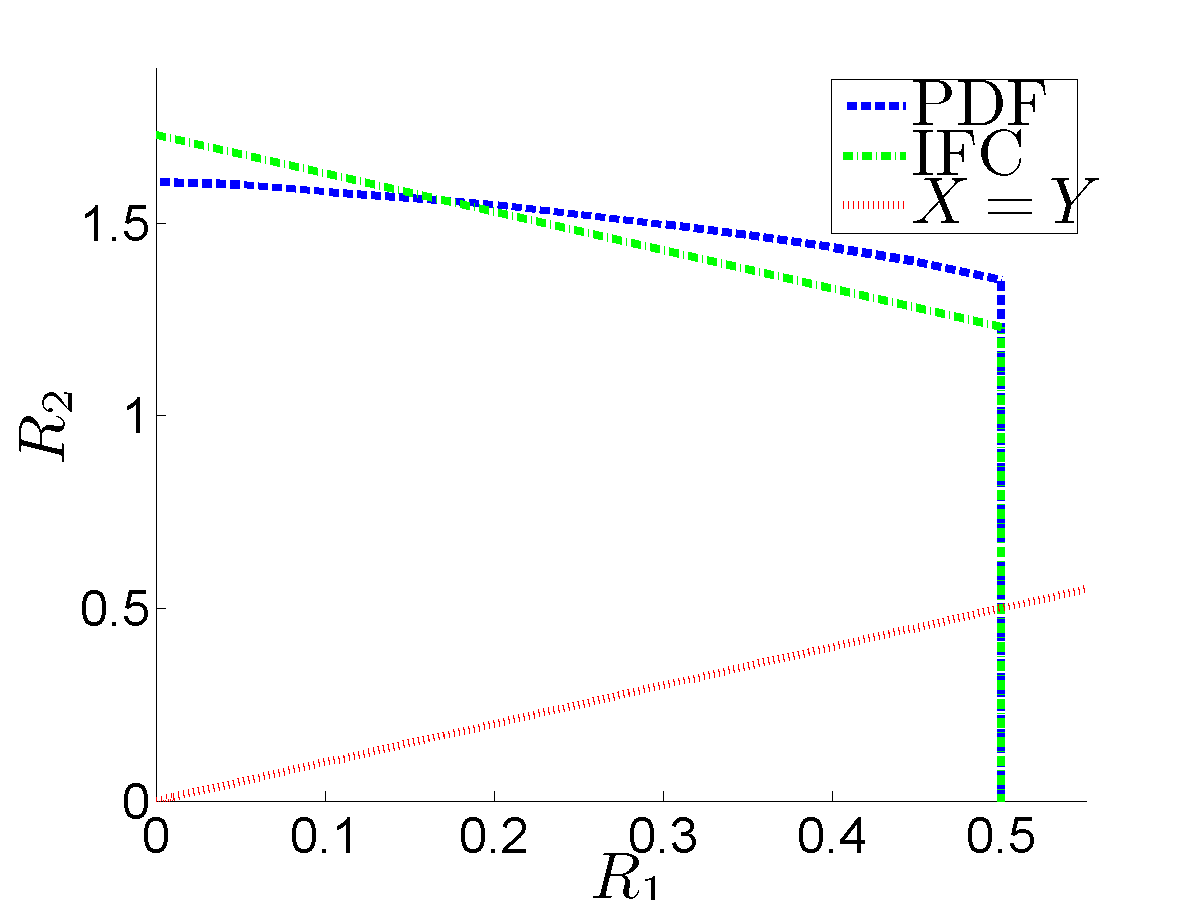}}
\subfigure[]{\label{fig:rate2}\includegraphics[width=0.5\columnwidth]{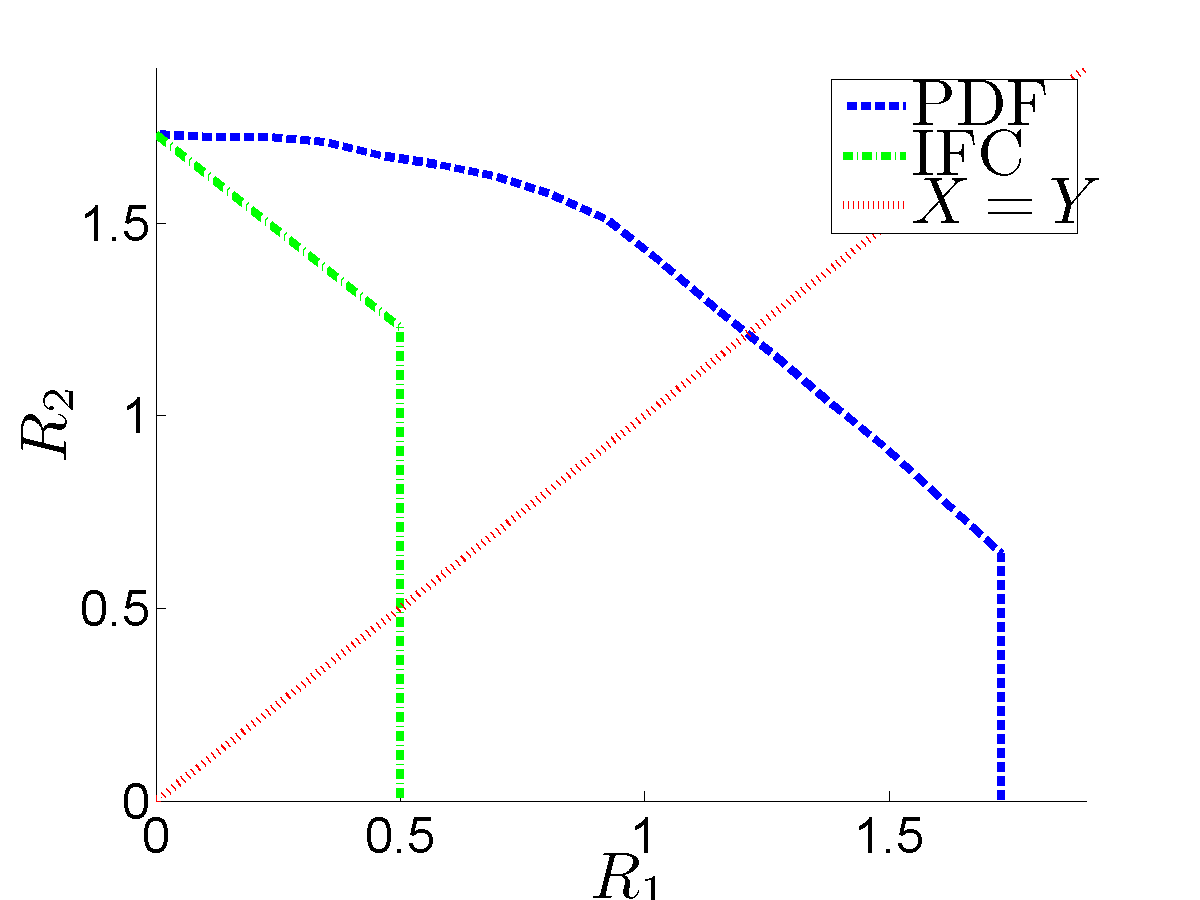}}
\subfigure[]{\label{fig:rate3}\includegraphics[width=0.5\columnwidth]{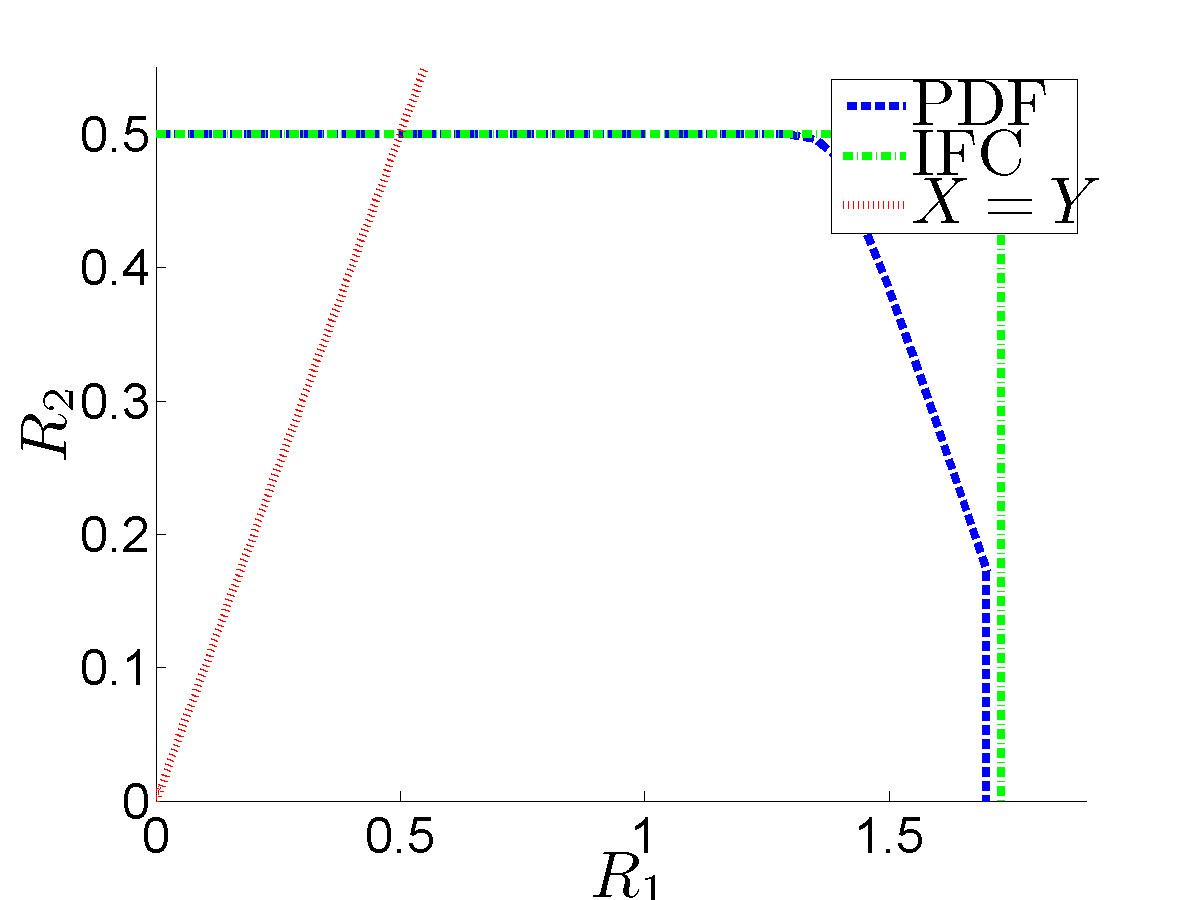}}
\subfigure[]{\label{fig:rate4}\includegraphics[width=0.5\columnwidth]{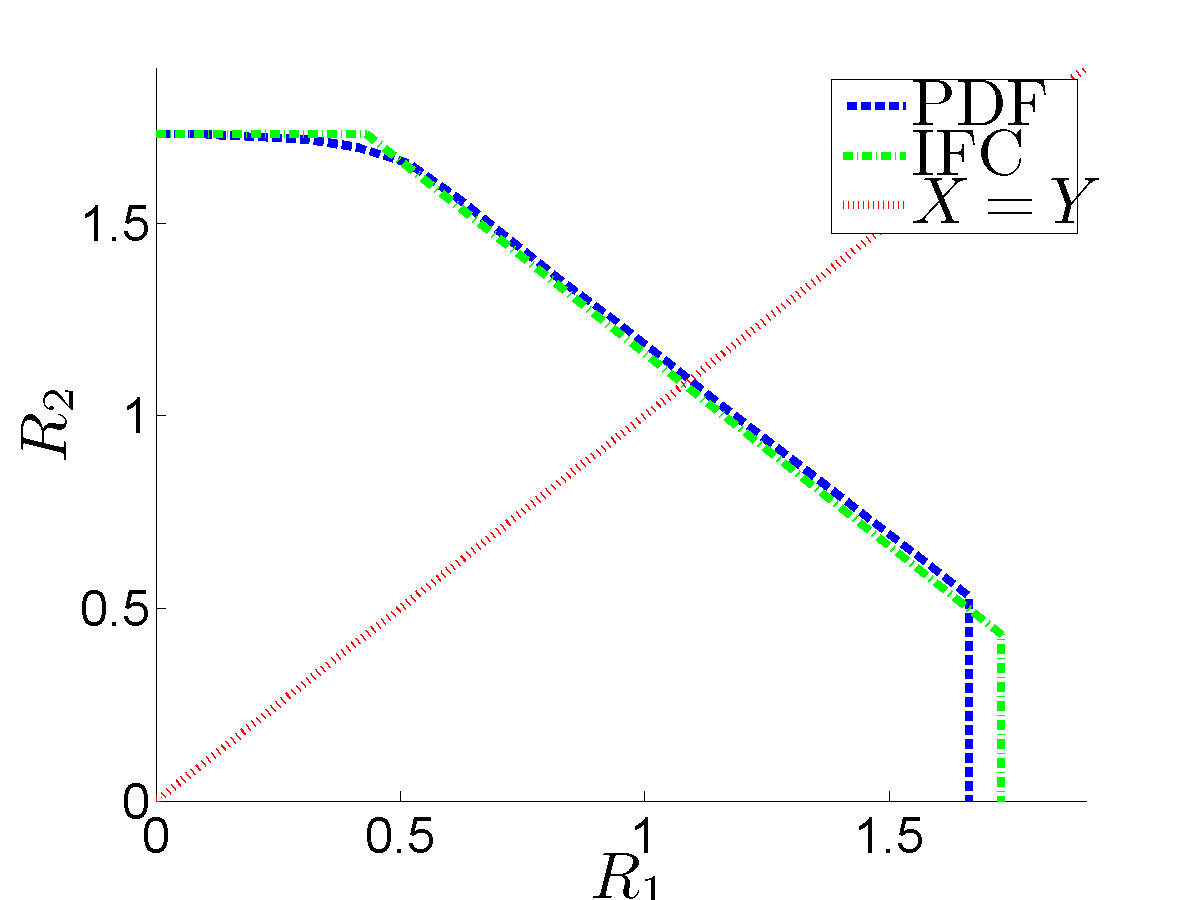}}
\subfigure[]{\label{fig:rate5}\includegraphics[width=0.5\columnwidth]{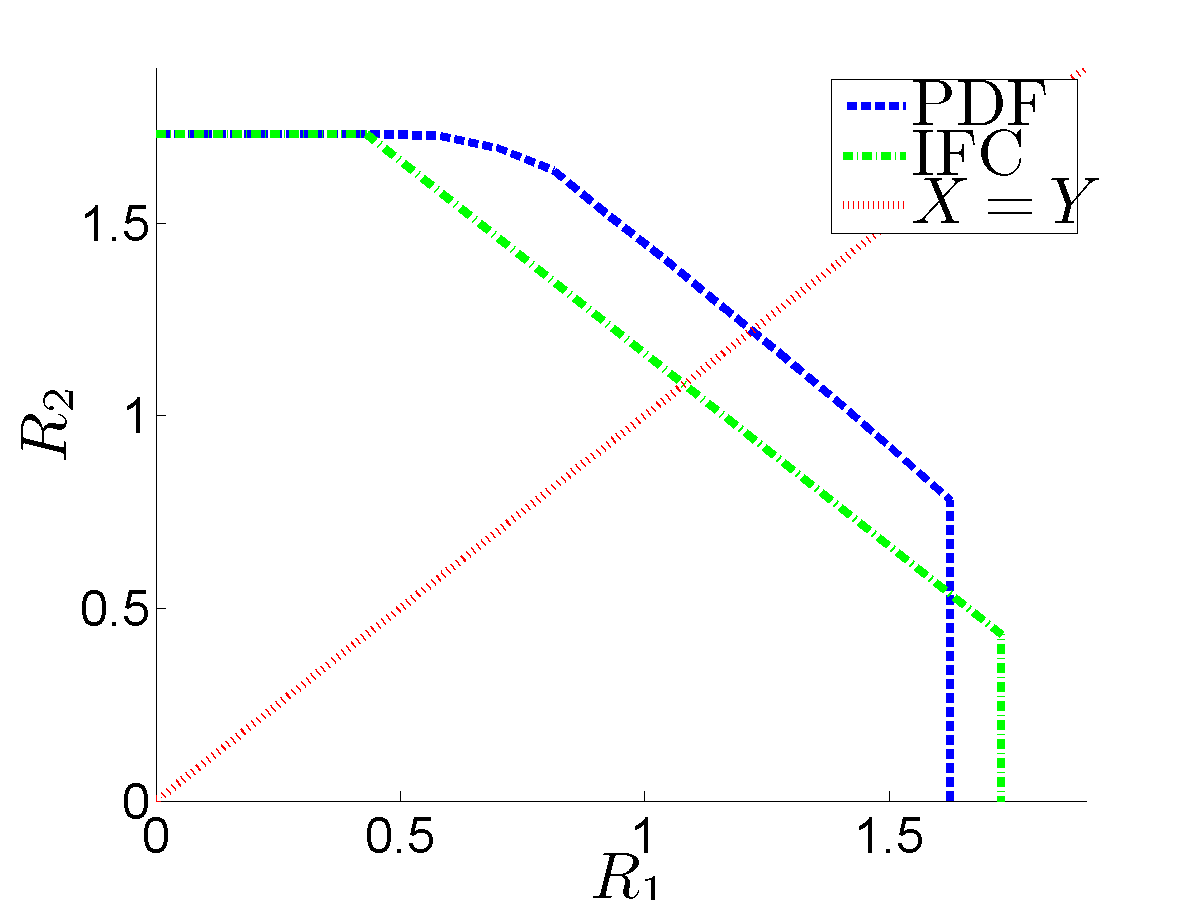}}
\end{center}
\caption{The achievable rate region for collaborative, partial decode and forward (PDF) and non collaborative, interference channel (IFC) in a pairwise collaborative network with different scenarios for channel conditions: a) $h_{12} = 1,  h_{13} = 10, h_{14} = 1,  h_{23} = 10,; h_{24} = 10$ and $h_{34} = 1$ b) $h_{12} = 10, h_{13} = 10, h_{14} = 1,  h_{23} = 10,; h_{24} = 10$ and $h_{34} = 10$ c) $h_{12} = 10, h_{13} = 10, h_{14} = 10, h_{23} = 1,;  h_{24} = 10$ and $h_{34} = 10$ d) $h_{12} = 1,  h_{13} = 10, h_{14} = 10, h_{23} = 10,; h_{24} = 10$ and $h_{34} = 1$ e) $h_{12} = 10, h_{13} = 10, h_{14} = 10, h_{23} = 10,; h_{24} = 10$ and $h_{34} = 10$.}
\label{fig:rates}
\end{figure}

Figure~\ref{fig:rates} demonstrates the trade off between the achieved rate region of the source and relay pairs. The achievable rate region of the source pair expands as the transmitter increases its transmit power. Similar to the source pair, increasing the transmit power of the relay pair increases the achievable rate of the relay pair.

First we investigate the case where the communication channel between the source pair is poor. The channel condition $h_{12} = 1,  h_{13} = 10, h_{14} = 1,  h_{23} = 10,; h_{24} = 10$ and $h_{34} = 1$ exemplifies such a condition. Figure~\ref{fig:rate1} shows the rate region of the involved pairs employing the proposed collaboration scheme in conjunction with rate region of interference channel. We observe that under this condition, collaboration offers small capacity gain. We also observe that the relay pair has incentive to collaborate and obtain more rate than interference channel, only if the source pair demands for more rate. We plotted the line $X = Y$ to investigate the achievable rate on condition that both pairs demands equal rates, i.e. $R_1 = R_2$. We observe that under this condition collaboration is not beneficial.

Exploiting strong communication links between the 1st and the 2nd users and between the 3rd and the 4th users, pairs obtain a considerable capacity gain which is shown in Figure~\ref{fig:rate2} with $h_{12} = 10, h_{13} = 10, h_{14} = 1,  h_{23} = 10,; h_{24} = 10$ and $h_{34} = 10$. In this scenario, the source pair is suffering from poor direct channel gain, between the 1st and the 4th users, however, the communication channel between pair users to relay users is strong. In that case collaboration enhances the achievable rate of both pairs. It also offers dramatic gain if both pairs are interested in equal rates.

Collaboration only improves the achievable rate if the direct link between relay pair, i.e. the 2nd and the 3rd user, is strong. Otherwise as shown in Figure~\ref{fig:rate3}, with $h_{12} = 10, h_{13} = 10, h_{14} = 10, h_{23} = 1,;  h_{24} = 10$ and $h_{34} = 10$, collaboration does not enlarge the rate region. However, increasing the channel gain between the 3rd and 4th users and between the 1st and the 2nd, the pairs gain as much as the non collaborative scheme (see Figure~\ref{fig:rate4}), with $h_{12} = 1,  h_{13} = 10, h_{14} = 10, h_{23} = 10,; h_{24} = 10$ and $h_{34} = 1$. This emphasizes that the efficiency of proposed scheme significantly depends on the channel gain between the relay users.

Lastly, in equal channel condition for both pairs, i.e. $h_{12} = 10, h_{13} = 10, h_{14} = 10, h_{23} = 10,; h_{24} = 10$ and $h_{34} = 10$ (Figure~\ref{fig:rate5}), the channel gain between the 1st and the 2nd users and between the 3rd and the 4th users, increases the capacity gain for involved pairs.

\section{Conclusion}
\label{sec:conclusion}
We have considered a network of collaborative transmitter-receiver pairs in which one pair (relay pair) acts as relay to assist the source pair in delivering the message of the source pair as well as its own private message. We have studied partial decode and forward collaborative schemes and established the capacity of this coding schemes for the PCN. In the proposed scheme we let the transmitter of the source pair to split its message into two independent parts. The relay pair decodes and forwards only one part of the message of the source pair and re-encodes and transmits the decoded message along with the relay pair private message. Having decoded both messages, the receiver of the relay pair decodes and transmits the message of the source pair to the intended destination. For AWGN PCNs, we have characterized the achievable rate regions. We have also provided numerical results and compared the proposed collaboration scheme with achievable rate of a non collaborative scheme, i.e. interference channel. We have examined the channel conditions under which such a collaboration is beneficial. We have shown that when the channel gain between the source pair is week collaboration offers capacity gain to both pairs. However, if the channels condition between the involved pairs are poor such a collaboration is not beneficial.


\begin{thebibliography}{10}
\providecommand{\url}[1]{#1}
\csname url@samestyle\endcsname
\providecommand{\newblock}{\relax}
\providecommand{\bibinfo}[2]{#2}
\providecommand{\BIBentrySTDinterwordspacing}{\spaceskip=0pt\relax}
\providecommand{\BIBentryALTinterwordstretchfactor}{4}
\providecommand{\BIBentryALTinterwordspacing}{\spaceskip=\fontdimen2\font plus
\BIBentryALTinterwordstretchfactor\fontdimen3\font minus
  \fontdimen4\font\relax}
\providecommand{\BIBforeignlanguage}[2]{{%
\expandafter\ifx\csname l@#1\endcsname\relax
\typeout{** WARNING: IEEEtran.bst: No hyphenation pattern has been}%
\typeout{** loaded for the language `#1'. Using the pattern for}%
\typeout{** the default language instead.}%
\else
\language=\csname l@#1\endcsname
\fi
#2}}
\providecommand{\BIBdecl}{\relax}
\BIBdecl

\bibitem{vandermeulen1971ttc}
E.~C. van~der Meulen, ``{Three-terminal communication channels},'' \emph{Adv.
  Appl. Prob}, vol.~3, no.~1, pp. 120--154, 1971.

\bibitem{cover1979ctr}
T.~Cover and A.~Gamal, ``{Capacity theorems for the relay channel},''
  \emph{{IEEE} Trans. Inf. Theory}, vol.~25, no.~5, pp. 572--584, 1979.

\bibitem{liang2006crb}
Y.~Liang and V.~V. Veeravalli, ``Cooperative relay broadcast channels,''
  \emph{{IEEE} Trans. Inf. Theory}, vol.~53, no.~3, pp. 900--928, March 2007.

\bibitem{liang2007rrr}
Y.~Liang and G.~Kramer, ``Capacity theorems for cooperative relay broadcast
  channels,'' \emph{in Proc. 40th Annual Conference on Information Sciences and
  Systems}, pp. 1719--1724, 22-24 March 2006.

\bibitem{tannious2007rcp}
R.~Tannious and A.~Nosratinia, ``{Relay Channel With Private Messages},''
  \emph{{IEEE} Trans. Inf. Theory}, vol.~53, no.~10, pp. 3777--3785, 2007.

\bibitem{kramer2005csa}
G.~Kramer, M.~Gastpar, and P.~Gupta, ``{Cooperative Strategies and Capacity
  Theorems for Relay Networks},'' \emph{{IEEE} Trans. Inf. Theory}, vol.~51,
  no.~9, pp. 3037--3063, 2005.

\bibitem{akhavanastaneh2007cgr}
S.~A.~Astaneh and S.~Gazor, ``{Joint Relay and Node Selection in Collaborative
  Networks},'' \emph{24th Biennial Symposium on Communications QBSC'08}, June
  2008.

\bibitem{reznik2005brc}
A.~Reznik, S.~Kulkarni, and S.~Verdu, ``Broadcast-relay channel: capacity
  region bounds,'' \emph{Information Theory, 2005. ISIT 2005. Proceedings.
  International Symposium on}, pp. 820--824, 4-9 Sept. 2005.

\bibitem{sendonaris2003ucd1}
A.~Sendonaris, E.~Erkip, and B.~Aazhang, ``{User cooperation diversity. Part I.
  System description},'' \emph{{IEEE} Trans. Commun.}, vol.~51, no.~11, pp.
  1927--1938, 2003.

\bibitem{sendonaris2003ucd2}
------, ``{User cooperation diversity. Part II. Implementation aspects and
  performance analysis},'' \emph{{IEEE} Trans. Commun.}, vol.~51, no.~11, pp.
  1939--1948, 2003.

\bibitem{laneman2004cdw}
J.~Laneman, D.~Tse, and G.~Wornell, ``{Cooperative diversity in wireless
  networks: Efficient protocols and outage behavior},'' \emph{{IEEE} Trans.
  Inf. Theory}, vol.~50, no.~12, pp. 3062--3080, 2004.

\bibitem{hostmadsen2004arr}
A.~Host-Madsen, ``{On the achievable rate for receiver cooperation in ad-hoc
  networks},'' \emph{ISIT 2004. Proceedings. International Symposium on
  Information Theory}.

\bibitem{hostmadsen2006cbc}
------, ``{Capacity Bounds for Cooperative Diversity},'' \emph{{IEEE} Trans.
  Inf. Theory}, vol.~52, no.~4, pp. 1522--1544, 2006.

\bibitem{ng2007cgt}
C.~Ng, N.~Jindal, A.~Goldsmith, and U.~Mitra, ``{Capacity Gain From
  Two-Transmitter and Two-Receiver Cooperation},'' \emph{{IEEE} Trans. Inf.
  Theory}, vol.~53, no.~10, pp. 3822--3827, 2007.

\bibitem{cover1980mac}
T.~Cover, A.~Gamal, and M.~Salehi, ``{Multiple access channels with arbitrarily
  correlated sources},'' \emph{{IEEE} Trans. Inf. Theory}, vol.~26, no.~6, pp.
  648--657, 1980.

\bibitem{liang2008mac}
Y.~Liang and H.~V. Poor, ``Multiple-access channels with confidential
  messages,'' \emph{{IEEE} Trans. Inf. Theory}, vol.~54, no.~3, pp. 976--1002,
  March 2008.

\bibitem{cover2006eit}
T.~Cover and J.~Thomas, \emph{{Elements of Information Theory}}.\hskip 1em plus
  0.5em minus 0.4em\relax Wiley-Interscience New York, 2006.

\bibitem{sason2004arr}
I.~Sason, ``{On achievable rate regions for the Gaussian interference
  channel},'' \emph{Information Theory, IEEE Transactions on}, vol.~50, no.~6,
  pp. 1345--1356, 2004.

\end{thebibliography}

\end{document}